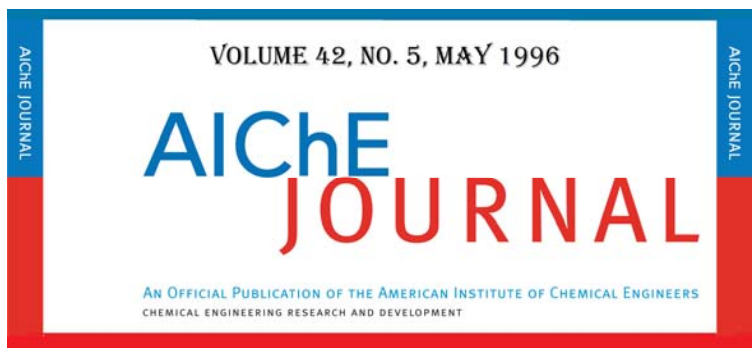

# Surface Tension Prediction for Pure Fluids

Joel Escobedo and G. Ali Mansoori*
University of Illinois at Chicago
(M/C 063) Chicago, IL 60607-7052

**Abstract**

In this paper we propose the following expression for surface tension of organic compounds:

$$\sigma = \left[\mathcal{P}(\rho_l - \rho_v)\right]^4$$

where

$$\mathcal{P} = \mathcal{P}_o \cdot (1-T_r)^{0.37} \cdot T_r \cdot \exp(0.30066/T_r + 0.86442 \cdot T_r^9)$$

In this equation $\rho_l$ and $\rho_v$ are the molar densities of liquid and vapor, respectively, $T_r = T/T_c$, $\mathcal{P}_o$ is a temperature-independent compound-dependent constant similar to the Sugden's parachor. This new expression, originally derived from the statistical-mechanics is shown to represent the experimental surface tension data of 94 different organic compounds within 1.05 AAD%. We also propose

$$\mathcal{P}_o = 39.6431 \cdot [0.22217 - 2.91042 \times 10^{-3} \cdot (\mathcal{R}^*/T_{br}^2)] \cdot T_c^{13/12}/P_c^{5/6}$$

as a corresponding states expression to correlate the temperature-independent parameter $\mathcal{P}_o$ for various compounds. In this equation $\mathcal{R}^* = \mathcal{R}_m/\mathcal{R}_{m,ref}$, $\mathcal{R}_m$ is the molar refraction, $\mathcal{R}_{m,ref}$ is the molar refraction of the reference fluid (methane), and $T_{br}$ is the reduced normal boiling point. When this generalized expression is used surface tensions for all the 94 compounds can be predicted within 2.57 AAD% for all temperatures investigated.

---

(*) Corresponding author, email: *Mansoori@uic.edu*

**Introduction**

One of the most striking demonstrations of the intermolecular forces is the tension at the surface of a liquid. At the molecular level, one may consider the boundary layer at the liquid-vapor interface to be a third phase with properties intermediate between those of a liquid and its vapor. From a qualitative point of view, a plausible explanation of this phenomenon would be that at the interface there are unequal forces acting upon the molecules. At low densities the molecules experience a sidewise attraction and toward the bulk liquid, however, they experience less attraction toward the bulk vapor phase. Therefore, there is always a tendency for the surface layer to minimize its area according to the total mass, constraints posed by the container, and external forces.

The surface tension ($\sigma$), generally employed as a quantitative index of this tension, is defined as the force exerted in the plane of the surface per unit length (*e.g.* dynes/cm). Numerous methods have been proposed to estimate the surface tension of pure liquids and liquid mixtures. An extensive revision of these methods is given by Hirschfelder, *et al.* (1964). One of the simplest is the empirical formula proposed by Macleod (1923). It expresses the surface tension of a liquid in equilibrium with its own vapor as a function of the liquid- and vapor-phase densities as:

$$\sigma = K(\rho_l - \rho_v)^4 \tag{1}$$

where K is a constant which is independent of temperature but is characteristic of the liquid under consideration. Sugden (1924) modified this expression as follows:

$$\sigma = \left[P \cdot (\rho^l - \rho^v)\right]^4 \tag{2}$$

where $P = K^{1/4}$, Sugden called this temperature-independent parameter (P) the parachor, and indicated a way to estimate it from molecular structure. Quayle (1953) used experimental surface tension and density data for numerous compounds to calculate the parachors of hydrocarbons. He was able to suggest an additive procedure to correlate P with structural contribution.

The right-hand side of Equation 2 implies that surface tension is very sensitive to the value of the parachor and liquid density. It has been shown that the parachor is a weak function of temperature for a variety of fluids and within wide ranges of temperature (Macleod, 1923; Sugden, 1924; Quayle, 1953), and thus it is generally assumed to be a constant. Equation 2 has been shown to be good for surface tension prediction, so long as experimental data for the parachor and equilibrium densities are employed. Thus, it may be considered as an equation of state for the interface (Boudh-Hir and Mansoori, 1990).

The good performance and extreme simplicity of its analytical form have made Equation 2 a very popular method for surface tension calculation (Weinaug and Katz, 1943; Lee and Chien, 1984; Hugill and van Welsenes, 1986; Gasem *et al.*, 1989; Fanchi, 1985, 1990; Ali, 1994). Nevertheless, there are various shortcomings on the use of this equation: (i) The parachor P is actually a temperature-dependent parameter whose functional form with temperature was not known, (ii) the empirical nature of the parachor poses difficulty in deriving a more accurate expression for it, and (iii) the absolute average percent deviation (AAD%) in surface tension prediction increases with increasing complexity of the molecular structure of the fluid under consideration. These observations are important both from the fundamental and practical point of view. Knowledge of the statistical-mechanical basis of Equation 1 may allow modifications to this expression for the purpose of improving its accuracy and/or extending its use to the calculation of interfacial tension.

**Development**

Macleod's empirical expression for surface tension calculation, as mentioned above, has proven to work very well for many substances and over a wide range of temperature. Nonetheless, deviations with respect to temperature are generally observed. Thus, efforts have been made to derive the functionality of surface tension with respect to temperature (van der Waals, 1894; Lovett, *et al.*, 1973; Guggenheim, 1945). The surface tension of Argon, nitrogen, and xenon has been measured over a very large range of temperature (Croxton, 1974). Cahn and Hilliard (1958) arrived at a power law expression for the temperature dependence of surface tension in the critical region. It should be pointed out, however, that a simple power law is not guaranteed to hold right for all temperatures down to the triple point. In fact, its applicability is confined to hold valid for the critical region (Levelt Sengers and Sengers, 1981) and for molecules of high symmetry (Croxton, 1974). Based on the principle of corresponding states, Brock and Bird (1955) developed an expression for nonpolar liquids utilizing the power law concept applicable to temperatures away from the critical point. The accuracy of their expression is similar to that of Equation 1. Hakim *et al.*(1971) modified Brock and Bird's expression to include polar liquids. However, the general reliability of their expression is not known since the values of the constants appearing in the equation are known only for a few substances. Sivaraman *et al.* (1984) developed a correlation for surface tension prediction of organic compounds within $0.45 < T_r < 0.97$. This correlation, however, is very sensitive to the values of the acentric factor, critical temperature, and critical pressure. Somayajulu (1988) has proposed a three-parameter generalized equation for surface tension from the triple point to the critical point. Although this empirical expression is very accurate, it involves three different compound-specific parameters which are known only for the compounds analyzed. Lopez-Perez *et al.* (1992) used the gradient theory to predict the surface tensions of nonpolar, qadrupolar, polar, and nonpolar. Their predictions are good, however, the absolute average percent deviations observed are rather large for all the compounds that they investigated.

In spite of the various efforts made to describe the functionality of surface tension with respect to temperature, there is still a need for an analytic expression which would be applicable over a larger range of temperature and valid for a variety of compounds.

From the statistical-mechanical basis of Macleod's formula it is feasible to modify this expression in order to increase its accuracy as well as the temperature range of its applicability.

There have been a number of efforts made to justify the success of Macleod's formula from theoretical basis (Fowler, 1937; Green, 1969; Henderson, 1980). These start from the classical thermodynamic expression which relates the surface tension and the surface internal energy (*i.e.* $u_s = S\{\gamma - T(\partial\gamma/\partial T)]$. However, the use of certain simplifying statistical-mechanical approximations (Green, 1969; Henderson, 1980) in these approaches leads to results different from the Macleod's equation. Thus, the temperature-dependence of surface tension is not easily observed.

Starting with the statistical-mechanical definition of the surface tension, Boudh-Hir and Mansoori (1990) have shown that this property, as a first approximation, is given by the Macleod formula. As a result, the law in power four of the difference in densities is obtained. However, they have shown that the constant, K, which depends on the nature of the fluid under consideration is not entirely independent of temperature. The only simplification to be made is to consider the particles to interact via a generalized additive pairwise potential. This interaction, however, is considered to depend on the position and orientations of the molecules (*i.e.* the particles are not spherical and the potential is not necessarily a radial function).

The statistical-mechanical expression for surface tension derived by Boudh-Hir and Mansoori (1990) is,

$$\sigma = \{(kT/4)\tau^{4-2g}(z/z_c)\,\zeta(\tau, \rho_l, \rho_v)\}(\rho_l - \rho_v)^4 \qquad (3)$$

where,

$$\zeta(\tau, \rho_l, \rho_v) = \int \partial_{z_1}\chi(1;\xi)e^{\rho_c\tau\chi(1;\xi)}\partial_{z_2}\chi(2;\xi)e^{\rho_c\tau\chi(2;\xi)}c(1,2)(r_{12}^2 - z_{12}^2)dz_1 d\Omega_1 dr_2 d\Omega_2 \qquad (3a)$$

In these equations, k is the Boltzmann constant; T is the temperature of the system under consideration; $\tau=(1-T_r)$; $g$ is an exponent; $z = (2\pi mkT/h^2)^{1/2} e^{\mu/kT}$ is the activity; the subscript c denotes the value of the activity at the critical temperature (*i.e.* $z_c = (2\pi mkT_c/h^2)^{1/2} e^{\mu/kT_c}$); $\mu$ is the chemical potential; h is the Plank's constant; $\rho_c\tau\chi(i;\xi) = \Delta c(i)$; $\rho_c$ is the critical density; $\Delta c(i) = [c(i)-c_c(i)]$ (*i.e.* the value of the one-particle direct correlation function at the temperature of interest minus the value of the one-particle direct correlation at the critical temperature). $\chi(i;\xi)$ is given by the expression:

$$\chi(i;\xi_0) = \int c(i,j;\xi)\left\{-g_c v\ (i,j)\ e^{-g_c w(i,j)} + g_c \mu - 3/2\right\} e^{\Delta c(j;\xi)} \tag{3b}$$

$c(i,j;\xi)$ is the two-particle direct correlation function; $w(i,j)$ is a pairwise potential; $v(i,j)$ is the mean force potential; $\xi$ is an order parameter which depends on $\tau$ (i.e. $\xi = 0$ corresponds to the system at its critical temperature, while $\xi = 1$ is associated with the system at the temperature of interest); $g_c$ is the value of the exponent $g$ at the critical temperature. Complete details on the derivation of Equation 3 are given elsewhere (Boud-Hir and Mansoori, 1990). Comparing Equations 2 and 3 it can be shown that,

$$\mathcal{P} = \mathcal{P}_o \cdot (1-T_r)^{1-g/2} \cdot T_r \cdot \text{Exp}(\mu_r/T_r) \cdot [\zeta(\tau, \rho^l, \rho^v)]^{1/4} \tag{4}$$

where $\mathcal{P}_o$ is a temperature-independent compound-specific constant; $g$, an adjustable parameter; and $\mu_r$ is the reduced chemical potential.

Considering the fact that for most organic substances we know little about the one- and two-particle direct correlation functions as well as the intermolecular potential it is almost impossible to calculate $\zeta$. Cosidering the success of the theory of parachors (Sugden, 1924) in correlating surface tension data, it can be safely assumed that $\zeta$ is a very weak function of densities. However, its temperature dependence is not known. As a result we write the expression for $\mathcal{P}$ in the following form,

$$\mathcal{P} = \mathcal{P}_o \cdot (1-T_r)^{1-g/2} \cdot T_r \cdot \text{Exp}(\theta_r/T_r); \quad \text{and} \quad \theta_r = \mu_r + (T_r \cdot \ln\zeta)/4 \tag{5}$$

by combining Equation 2 and 5 the following expression for the surface tension is obtained,

$$\sigma = \left[\mathcal{P}_o \cdot (1-T_r)^{1-g/2} \cdot T_r \cdot \text{Exp}(\theta_r/T_r) \cdot (\rho^l - \rho^v)\right]^4 \tag{6}$$

where $\rho^l$ and $\rho^v$ are the equilibrium molar densities of liquid and vapor phases, respectively. The temperature-dependent property, $\theta_r$, may be assumed to have the following simple form:

$$\theta_r = a_o + a_1 \cdot T_r^n \tag{7}$$

where $a_o$, $a_1$, and $n$ may be regarded as universal constants.

The temperature correction term in Equation 6 thus derived should be valid for all temperatures and for all compounds.

In order to evaluate the constants in Equations 6 and 7 reliable experimental data over a wide range of temperature are needed. Recently, Grigoryev, *et al.*(1992) have reported experimental surface tension data for n-Pentane, n-Hexane, n-Heptane, and n-Octane which cover the entire temperature range from the triple point to the critical point. Thus, these data were considered to be appropriate. After analysis of these data it was concluded that the most complete sets of data were those of n-Hexane and n-Heptane for they cover the entire temperature range more completely. From Equation 6 it is noticed that data for the equilibrium densities are also needed for the analysis to be completed. Considering the fact that for most substances experimental data for equilibrium densities may not be readily available, we decided to use an accurate equation of state for the purpose of evaluating the constants $a_0$, $a_1$, and $n$ appearing in Equation 7. This is important since in this way we eliminate the need for experimental equilibrium density data.

**Calculation of Equilibrium Densities**

Recently, Riazi and Mansoori (1993) proposed a simple equation of state (R-M EOS) that accurately predicts fluid densities. This equation is a modification of that originally proposed by Redlich and Kwong, (1949) such that,

$$P = \rho RT/(1-b\rho) - a \rho^2/[T^{1/2} \cdot (1+b\rho)]; \quad b = (0.08664 \, RT_c/P_c) \cdot \delta(\mathcal{R}^*, T/T_c) \tag{8}$$

These authors consider the fact that for liquid systems, in which the free space between the molecules decreases, the role of parameter b becomes more important than that of parameter a. For this reason, "a" is considered to be constant and given by: $a = 0.42748 RT_c/P_c$. Parameter b, however, was modified using the molecular theories of perturbations and refractive index such that $b = (0.08664 \, RT_c/P_c) \cdot \delta(\mathcal{R}^*, T_r)$; R is the universal gas constant; $T_c$ and $P_c$ are the critical temperature and pressure respectively; $\delta$ is a temperature-dependent parameter given by:

$$\delta^{-1} = 1 + \{0.02[1 - 0.92 \cdot \exp(-1000 \, |T_r - 1|)] - 0.035(T_r - 1)\}(\mathcal{R}^* - 1); \quad \mathcal{R}^* = \mathcal{R}_m/\mathcal{R}_{m,ref}$$

$\mathcal{R}^*$ is a dimensionless molar refraction; $\mathcal{R}_m$ is the molar refraction and $\mathcal{R}_{m,ref}$ is the molar refraction of a reference fluid which in this case is methane with a value of 6.987. $\mathcal{R}_m$ may be calculated using the Lorentz-Lorenz function:

$$\mathcal{R}_m = (4\pi N_A/3)[\alpha + \mu^2 \cdot f(T)] = (1/\rho)(n^2-1)/(n^2+2)$$

$N_A$ is the Avogadro's number; $\alpha$ is the polarizability; $\mu$ is the dipole moment; $\rho$ is the molar density; and n is the sodium-D light refractive index of a liquid at 20°C and 1 atm. In this derivation, it is considered the fact that the molar refraction is, as a first approximation, independent of temperature. Furthermore, $\mathcal{R}^*$ is even less sensitive to temperature.

This equation was selected for density calculation and compared its predictions against the available experimental data (Hall, 1986) for the compounds of interest. These results are reported in Table 1 along with the properties of 94 organic compounds. From this table it can seen that density predictions using this equation of state are good for most compounds except for those whose molecular structure are highly assymetric and/or are polar.

The necessary vapor pressures are calculated using a recently proposed (Edalat and Mansoori, 1993) generalized vapor pressure equation for various fluids. This new expression is as follows,

$$\ln (P/P_c) = [a(\omega).\tau + b(\omega).\tau^{1.5} + c(\omega).\tau^3 + d(\omega).\tau^6]/(1 - \tau) \tag{9}$$

In this equation,

$a(\omega) = -6.1559 - 4.0855.\omega;$　　$b(\omega) = 1.5737 - 1.0540.\omega - 4.4365 \times 10^3.d(\omega);$
$c(\omega) = 0.8747 - 7.8874.\omega;$　　$d(\omega) = (-0.4893 - 0.9912.\omega + 3.1551.\omega^2)-1;$　　$\tau = 1- T_r$

$P_s$ is the saturation pressure, and $\omega$ is the acentric factor. In Table 1 shows the absolute average percent deviation (AAD%) in the calculation of the saturation pressure at the normal boiling for all 94 compounds. It is observed that the saturation pressure at the normal boiling point is represented within an overall AAD% of 1.76.

**Results and Discussion**

Equilibrium densities for n-Hexane and n-Heptane were calculated by predicting first the saturation pressure from Equation 9 and then using Equation 8 to predict the vapor- and liquid-phase densities. These were used along with the experimental surface tension data of Grigoryev, *et al.*, (1992) to evaluate the universal constants appearing in Equation 7. The values for the constants thus obtained are: $a^0 = 0.30066$, $a^1 = 0.86442$, and $n = 10$. By introducing Equation 7 into Equation 6 using the constants just found the following new expression for the surface tension of a fluid was obtained,

$$\sigma = \left[ \mathcal{P}\text{b} \cdot (1-T_r)^{0.37} \cdot T_r \cdot \exp\left(0.30066/T_r + 0.86442 \cdot T_r^9\right) \cdot (\rho^l - \rho^v) \right]^4 \tag{10}$$

In this expression $\mathcal{P}$b is a temperature-independent constant characteristic of the fluid under consideration similar to the Sugden's parachor.

      Experimental surface tension data (Jasper, 1972; Hall, 1986; Beaton and Hewitt, 1989) for 94 compounds and equilibrium densities calculated as previously explained were used to evaluate the constant $\mathcal{P}$b for these substances. The results obtained are reported in Table 1 along with the surface tension predictions using this new parameter $\mathcal{P}$b. The temperature range used in the analysis is also reported. For comparison purposes, Table 1 also shows the surface tension predictions obtained by using the experimental parachor (Quayle, 1953) and the equilibrium densities calculated using the R-M EOS. It is noticed from this table an overall AAD% of 1.05 in surface tension prediction for all 94 compounds when the new parameter $\mathcal{P}$b is used along with Equation 10. This proves this equation to be valid for a large variety of organic compounds. It can also be noticed from this table an overall average absolute deviation of 16.81% when the experimental parachor is used along with Equation 2.

      In order to show the predicting capabilities of equation 10, the results obtained for 12 different compounds are plotted in Figures 1-3. Figure 1 shows the results obtained for methane, Ethane, propane and n-butane over a wide range of temperature. These are compared against the experimental surface tension data (Jasper, 1972; Hall, 1986; Beaton and Hewitt, 1989) and against the predictions obtained using the experimental parachor and Equation 2 (equilibrium densities were calculated as explained above). From this figure it can be seen that predictions by Equation 10 are quite good. Figure 2 present sthe results obtained for n-Pentane, n-Hexane, n-Heptane, and n-Octane over temperatures ranging from the triple point to near the critical point. These results are compared with the experimental data of Grigoryev, *et al.*, (1992) and with the predictions obtained using the experimental parachor in Equation 2 (equilibrium densities were calculated as explained above). Notice from this figure that the experimental data for all four compounds are represented quite well by Equation 10 except in the region close to the critical point. This is because the R-M EOS, as all other equations of state, does not perform well around the critical region. Figure 3 depicts the results for Toluene, Cyclopentane, Ethylbenzene, and Carbon Tetrachloride over a wide temperature range. The experimental data for these compounds were obtained from Jasper (1972), Hall (1986) and Beaton and Hewitt (1989). It can also be noticed from this figure that the experimental data is well represented by Equation 10. Judging from Figures 1-3 we may say that overall performance of our method for surface tension prediction is goog. Knowing that parameter $\mathcal{P}$b provides such a good surface tension prediction for all the compounds and for all temperatures investigated, it is only logical to attempt to find a way to correlate this parameter using the corresponding states principle.

Following the principle of corresponding states (Hirschfelder, 1964) a reduced surface tension may be defined and expected to be a universal function of the reduced temperature, as follows,

$$\sigma_r = \sigma / \left[ P_c^{2/3} \cdot (kT_c)^{1/3} \right] \tag{11}$$

From Equations 10 and 11 it can be shown that a reduced parameter $P_r$ can be expressed as follows,

$$P_r = P_o \cdot P_c^{5/6} / \left( 39.6439 \, T_c^{13/12} \right) \tag{12}$$

According to the principle of corresponding states, it is expected that a reduced parameter thus defined would be a universal function of the acentric factor. The calculated values of $P_r$ for all 94 compounds listed in Table 1 were plotted against the acentric factor in Figure 4. From this figure one may notice a trend as a function of acentric factor. However, the scattering of the data is rather large. Therefore, we concluded that attempts to correlate our reduced parameter ($P_r$) to acentric factor would not be a feasible approach.

From the Lorentz-Lorenz function for molar refraction ($R_m$) defined earlier in this paper, one may notice that this quantity ($R_m$) depends on the polarizability of the molecule and the dipole moment. Furthermore, the molar refraction provides an approximate measure of the actual volume (without free space) of the molecules per unit mole (Hirschfelder, 1964). Therefore it implicitly accounts for the asymmetry of the molecules. This fact gave the authors confidence that $R_m$ would be useful in finding a good correlation for $P_r$. It has also been found that substances with higher polarities have higher viscosities, normal boiling and freezing points. Therefore, $R_m$ and the normal-boiling-point temperature could be useful for the purposes at hand.

In Figure 2 the reduced parachor $P_r$ (value at the normal boiling point, $T_b$) versus $R^*/T_{br}^2$ have been plotted. $R^*$ as previously defined and $T_{br}$ is the reduced normal boiling point temperature. Judging from Figure 5 it may be said that there is a clear relationship between $P_r$ and $R^*/T_{br}^2$. Thus, the following simple correlation for $P_r$ was obtained,

$$P_r = 0.22217 - 2.91042 \times 10^{-3} \cdot ( R^*/T_{br}^2 ) \tag{13}$$

Therefore, the new expression proposed to predict parameter $P_o$ in equation 10 is the following,

$$P_o = 39.6431 \cdot [0.22217 - 2.91042 \times 10^{-3} \cdot ( R^*/T_{br}^2 )] \cdot T_c^{13/12} / P_c^{5/6} \tag{14}$$

Equations 10 and 14 were used along with the Riazi-Mansoori equation to predict the surface tension for the 94 compounds of interest. These results are reported in Table 1 from which it can seen that the predictions obtained from these new equations are quite good. The experimental surface tension data for all 94 compounds can be represented within an absolute average deviation (AAD%) of 2.57% for all temperatures investigated.

Hugill and van Welsenes, (1986) proposed the following correlation for the prediction of the Sugden's parachor:

$$P = 40.1684 \cdot (0.151 - 0.0464 \cdot \omega) \cdot T_c^{13/12} / \mathcal{P}_c^{5/6}. \tag{15}$$

For comparison purposes this correlation was used to predict the surface tension for the compounds of interest using Equation 2 and calculating the densities as explained above. These results are also reported in Table 1. From this table it may be noticed an overall absolute average deviation of 16.75% for all temperatures investigated for this correlation.

Table 2 contains the comparisons made between the experimental surface tension data and the values predicted by the present method and by other methods (Macleod, 1923; Brock and Bird, 1955; Sivaraman, 1984). This table shows that the method proposed in this paper for surface tension calculation performs better than the other methods. It can also be noticed that it performs equally well for all representative compounds (*i.e.* low-boiling-point, linear and branched alkanes, cyclic and branched cyclic, aromatic and alkyl-substituted aromatic, as well as halogenated compounds). It should be pointed out that the method proposed by Sivaram *et al.* (1984) was found to be very sensitive to the values of acentric factor employed and to a lesser extent to the values of the critical pressure and temperature. On the other hand, the method proposed in this work is not very sensitive to the value of these physical properties.

**CONCLUSIONS**

This paper presents a new expression for surface tension which contains a temperature correction term derived from statistical mechanics. It also introduces a corresponding-states correlation to predict the parameter $\mathcal{P}_b$ in this new equation as a function of molar refraction and normal boiling point temperature. This represents an accurate and generalized expression to predict surface tensions of pure fluids of industrial interest.

**Acknowledgement**

This research is supported by the National Science Foundation, Grant No. CTS-9108395.

**Notation**

| | |
|---|---|
| a | energy parameter in equation of state |
| b | volume parameter in equation of state |
| c($i$) | one-particle direct correlation function at the temperature of interest |
| c($i,j$) | two-particle direct correlation function |
| g | an exponent related to the difference in densities |
| k | Boltzmann constant |
| K | constant in Equation 1 |
| n | sodium-D light refractive index of a liquid at 20°C and 1 atm |
| P | Sugden's Parachor |
| $P_c$ | critical pressure |
| $\mathcal{P}b$ | compound-characteristic constant similar to the Sugden's parachor |
| $\mathcal{P}r$ | reduced parameter $\mathcal{P}b$ |
| $T_b$ | normal-boiling-point temperature |
| $T_c$ | critical temperature |
| r$ij$ | separation distance between particles $i$ and $j$ |
| **r**$ij$ | vector joining the center of masses of particles $i$ and $j$ |
| $\mathcal{R}^*$ | dimensionless molar refraction |
| $\mathcal{R}_m$ | molar refraction |
| $\mathcal{R}_{m,ref}$ | molar refraction of the reference fluid (methane) |
| v($i,j$) | mean force potential |
| w($i,j$) | pairwise potential |
| z | activity |

*Greek letters*

| | |
|---|---|
| $\alpha$ | polarizability of the molecules |
| $\gamma$ | critical exponent for surface tension |
| $\delta$ | temperature-dependent parameter in equation of state |
| $\theta_r$ | temperature dependent parameter |
| $\mu$ | dipole moment |
| $\mu$ | chemical potential |
| $\xi$ | order parameter |
| $\rho$ | density |
| $\rho_c$ | critical density |
| $\sigma$ | surface tension |
| $\Omega_i$ | orientation of particle $i$ |
| $\omega$ | acentric factor |

*Subscripts*

| | |
|---|---|
| c | property evaluated at the critical temperature |
| l | liquid |
| v | vapor |
| r | reduced property |

Table 1. Properties of organic compounds and comparisons against experimental data for density and surface tension over the temperature ranges indicated. The comparisons of saturation pressure at the normal boiling point are also shown in this table.

| Compound | $T_c$ (°K) | $P_c$ (bar) | $\mathcal{R}^*$ | $\omega$ | $T_b$ (°K) | $P$ | $\mathcal{P}^0$ | Temp. Range (°K) | AAD% density | AAD% $P_{sat}(T_b)$ $\mathcal{P}^0$ Eq10 | AAD%, Surface Tension Prediction using | | |
|---|---|---|---|---|---|---|---|---|---|---|---|---|---|
| | | | | | | | | | | | Exp $P$ Eq.2 | Calc $\mathcal{P}^0$ Eqs10,14 | Calc $P$ Eqs.2,15 |
| Methane | 190.4 | 46. | 1.0 | 0.011 | 111.6 | 72.6 | 102.31 | 90 - 170 | 0.90 | 1.74 | 3.39 | 5.52 | 3.96 | 7.57 |
| Ethane | 305.4 | 48.8 | 1.620 | 0.099 | 184.6 | 110.5 | 158.50 | 113 - 280 | 1.10 | 0.87 | 3.18 | 11.54 | 4.00 | 8.16 |
| Propane | 369.6 | 42.5 | 2.259 | 0.153 | 231.1 | 150.8 | 215.23 | 143 - 330.7 | 1.40 | 1.06 | 4.38 | 10.05 | 4.52 | 8.21 |
| n-Butane | 425.2 | 38. | 2.929 | 0.199 | 272.7 | 190.3 | 270.21 | 173.- 385 | 1.10 | 0.50 | 1.86 | 6.72 | 6.72 | 6.23 |
| Ethylene | 282.4 | 50.4 | 1.504 | 0.089 | 169.3 | 100.2 | 143.20 | 113 - 173 | 1.30 | 0.09 | 1.64 | 4.05 | 1.88 | 1.87 |
| Iso-butane | 408.2 | 36.5 | 2.955 | 0.183 | 261.4 | * | 266.81 | 173 - 283 | 1.40 | 0.05 | 0.60 | ** | 1.90 | 7.41 |
| n-pentane | 469.7 | 33.7 | 3.616 | 0.251 | 309.2 | 231.5 | 327.75 | 156 - 440 | 1.22 | 0.11 | 2.30 | 4.53 | 2.33 | 5.68 |
| Iso-Pentane | 460.4 | 33.9 | 3.620 | 0.227 | 301.0 | 230.0 | 318.29 | 273- 313 | 1.50 | 0.35 | 0.67 | 5.98 | 0.93 | 5.62 |
| n-Hexane | 507.4 | 30.1 | 4.281 | 0.299 | 341.9 | 270.4 | 387.60 | 175.12 - 450 | 1.46 | 0.16 | 1.55 | 7.30 | 2.84 | 4.26 |
| 2-Methylpentane | 497.5 | 30.1 | 4.286 | 0.278 | 333.4 | 270.0 | 376.96 | 273 - 343 | 0.57 | 0.11 | 0.84 | 2.60 | 0.82 | 4.61 |
| 3-Methylpentane | 504.5 | 31.2 | 4.265 | 0.272 | 336.4 | 267.7 | 370.33 | 273 - 343 | 1.64 | 0.19 | 0.26 | 6.57 | 0.83 | 6.84 |
| 2,2-Dimethylbutane | 488.8 | 30.8 | 4.289 | 0.232 | 322.8 | 266.4 | 358.62 | 273 - 303 | 4.83 | 0.26 | 0.62 | 18.79 | 2.95 | 16.60 |
| 2,3-Dimethylbutane | 500. | 31.3 | 4.267 | 0.247 | 331.1 | 266.2 | 363.54 | 273 - 333 | 2.93 | 0.12 | 0.51 | 12.24 | 2.52 | 13.25 |
| n-Heptane | 540.3 | 27.4 | 4.945 | 0.349 | 371.6 | 310.8 | 441.60 | 183.21 - 508 | 0.49 | 0.03 | 3.87 | 4.99 | 3.74 | 4.84 |
| 2-Methylhexane | 530.4 | 27.3 | 4.951 | 0.329 | 363.2 | 309.2 | 431.63 | 273 - 333 | 0.83 | 0.11 | 0.42 | 3.18 | 0.40 | 4.32 |
| 3-Methylhexane | 535.3 | 28.1 | 4.932 | 0.323 | 365.0 | 307.4 | 424.71 | 283 - 313 | 2.10 | 0.31 | 0.31 | 7.64 | 0.57 | 6.14 |
| n-Octane | 568.8 | 24.9 | 5.608 | 0.398 | 398.8 | 351.2 | 500.25 | 218.15 - 520 | 0.49 | 0.06 | 2.86 | 6.11 | 3.81 | 3.84 |
| 2-Methylheptane | 559.6 | 24.8 | 5.614 | 0.378 | 390.8 | 348.8 | 487.74 | 273 - 333 | 0.84 | 0.11 | 0.27 | 3.16 | 0.45 | 4.58 |
| 3-Methylheptane | 563.7 | 25.5 | 5.596 | 0.370 | 392.1 | 347.7 | 480.19 | 273 - 333 | 1.99 | 0.17 | 0.18 | 8.53 | 0.27 | 5.99 |
| 4-methylheptane | 561.7 | 25.4 | 5.599 | 0.371 | 390.9 | 347.4 | 479.88 | 273 - 333 | 2.01 | 0.08 | 0.17 | 8.38 | 0.27 | 5.82 |
| n-Nonane | 594.6 | 22.9 | 6.274 | 0.445 | 424.0 | 391.1 | 551.20 | 273 - 343 | 0.14 | 0.30 | 0.59 | 0.86 | 0.95 | 0.50 |
| Cyclopentane | 511.7 | 45.1 | 3.310 | 0.196 | 322.4 | 205.0 | 284.50 | 273 - 470 | 1.81 | 0.04 | 3.66 | 8.82 | 4.70 | 8.96 |
| Methylcyclopentane | 532.7 | 37.8 | 3.984 | 0.231 | 345.0 | 242.8 | 336.09 | 273 - 343 | 2.14 | 0.12 | 0.31 | 6.76 | 0.29 | 11.54 |
| Cyclohexane | 553.5 | 40.7 | 3.966 | 0.212 | 353.8 | 241.7 | 330.37 | 273 - 343 | 3.81 | 0.33 | 0.30 | 12.82 | 2.40 | 13.56 |
| 1,1-Dimethylcyclopentane | 547. | 34.4 | 4.648 | 0.273 | 361.0 | 281.2 | 364.11 | 273 - 343 | + | 1.70 | 0.21 | 39.90 | 3.82 | 17.96 |
| Methylcyclohexane | 572.2 | 34.7 | 4.652 | 0.236 | 374.1 | 281.3 | 379.74 | 273 - 343 | 4.55 | 0.01 | 1.27 | 19.14 | 2.36 | 24.45 |
| Ethylcyclopentane | 569.4 | 34. | 4.637 | 0.271 | 376.6 | 283.3 | 387.56 | 273 - 343 | 3.06 | 0.00 | 0.23 | 12.90 | 0.54 | 14.64 |
| 1,1-Dimethylcyclohexane | 591. | 29.6 | 5.302 | 0.238 | 392.7 | 318.8 | 446.73 | 273 - 333 | + | 0.40 | 0.43 | 3.36 | 2.52 | 27.56 |
| 1,2-Dimethylcyclohexane Cis | 606. | 29.6 | 5.273 | 0.236 | 402.9 | 317.4 | 462.23 | 273 - 333 | + | 1.57 | 0.90 | 11.04 | 4.73 | 24.87 |
| 1,2-Dimethylcyclohexane Trans | 596. | 29.6 | 5.314 | 0.242 | 396.6 | 320.3 | 448.75 | 273 - 333 | + | 0.78 | 0.33 | 3.58 | 0.66 | 29.43 |
| 1,3-Dimethylcyclohexane Cis | 591. | 29.6 | 5.338 | 0.224 | 393.3 | 321.3 | 441.06 | 273 - 333 | + | 6.08 | 0.62 | 12.24 | 2.31 | 36.76 |
| 1,3-Dimethylcyclohexane Trans | 598. | 29.7 | 5.297 | 0.189 | 397.6 | 318.7 | 451.19 | 273 - 333 | + | 16.38 | 0.68 | 0.60 | 2.35 | 36.32 |
| 1,4-Dimethylcyclohexane Cis | 598. | 29.7 | 5.224 | 0.234 | 397.5 | 318.8 | 451.56 | 273 - 333 | + | 2.28 | 0.42 | 0.83 | 1.73 | 28.07 |
| 1,4-Dimethylcyclohexane Trans | 587.7 | 29.7 | 5.339 | 0.242 | 392.5 | 322.7 | 437.27 | 273 - 333 | + | 4.99 | 0.75 | 18.11 | 2.72 | 33.28 |
| Ethylcyclohexane | 609.0 | 30. | 5.297 | 0.243 | 404.9 | 320.6 | 456.63 | 273 - 343 | 0.85 | 0.95 | 1.12 | 2.94 | 2.60 | 26.69 |
| Benzene | 562.2 | 48.9 | 3.748 | 0.212 | 353.2 | 206.14 | 289.59 | 283 - 343 | 0.20 | 0.43 | 0.55 | 1.42 | 2.77 | 11.87 |
| Toluene | 591.8 | 41. | 4.450 | 0.263 | 383.8 | 245.9 | 348.29 | 273 - 550 | + | 0.17 | 2.20 | 6.13 | 3.34 | 11.59 |
| o-Xylene | 630.3 | 37.3 | 5.124 | 0.310 | 417.6 | 283.3 | 394.89 | 273 - 343 | 1.26 | 0.02 | 0.62 | 6.39 | 0.65 | 16.90 |
| m-Xylene | 617.1 | 35.4 | 5.147 | 0.325 | 412.3 | 284.3 | 404.15 | 273- 343 | 0.56 | 0.11 | 0.87 | 1.98 | 0.94 | 13.00 |
| p-Xylene | 616.2 | 35.1 | 5.153 | 0.320 | 411.5 | 283.8 | 405.92 | 293 - 343 | 0.75 | 0.12 | 0.70 | 4.37 | 0.71 | 14.27 |
| Ethylbenzene | 617.2 | 36. | 5.118 | 0.302 | 409.3 | 284.3 | 398.69 | 273 - 593 | 0.70 | 0.26 | 1.18 | 6.83 | 1.43 | 12.46 |
| 1,2,3-Trimethylbenzene | 664.5 | 34.5 | 5.790 | 0.366 | 449.3 | 317.8 | 437.30 | 273 - 373 | 2.81 | 0.32 | 0.88 | 12.09 | 1.63 | 17.46 |
| 1,2,4-Trimethylbenzene | 649.2 | 32.3 | 5.824 | 0.376 | 442.5 | 320.4 | 452.94 | 273 - 343 | 0.27 | 0.36 | 0.58 | 1.02 | 0.58 | 13.78 |
| 1,3,5-Trimethylbenzene | 627.3 | 31.3 | 5.842 | 0.399 | 437.9 | * | 457.05 | 273 - 343 | 0.23 | 0.11 | 0.74 | ** | 0.74 | 8.63 |
| n-Decane | 617.7 | 21.2 | 6.915 | 0.489 | 447.3 | 431.15 | 602.86 | 273 - 393 | 0.74 | 1.06 | 1.28 | 3.65 | 1.28 | 0.55 |
| n-Undecane | 638.8 | 19.7 | 7.962 | 0.535 | 469.1 | 470.5 | 646.66 | 273 - 393 | 2.76 | 1.11 | 1.31 | 11.77 | 1.37 | 4.30 |
| n-Tridecane | 676. | 17.2 | 8.935 | 0.619 | 508.6 | 550.55 | 757.61 | 283 - 393 | 2.80 | 0.40 | 1.32 | 11.62 | 1.31 | 1.34 |
| 2,2-Dimethylhexane | 549.9 | 25.3 | 5.618 | 0.338 | 380.0 | 346.05 | 467.32 | 273 - 323 | 4.50 | 0.01 | 0.83 | 17.83 | 1.88 | 13.02 |

Table 1. Continued...

| Compound | Tc (°K) | Pc (bar) | $\mathcal{R}^*$ | ω | Tb (°K) | P | $\mathcal{P}^0$ | Temp. Range (°K) | AAD% density | AAD% Psat (Tb) | AAD%, Surface Tension Prediction using | | | |
|---|---|---|---|---|---|---|---|---|---|---|---|---|---|---|
| | | | | | | | | | | | $\mathcal{P}^0$ Eq10 | Exp P Eq.2 | Calc $\mathcal{P}^0$ Eqs10,14 | Calc P Eqs.2,15 |
| 2,4-DimethylHexane | 553.5 | 25.6 | 5.600 | 0.343 | 382.6 | 345.2 | 466.63 | 273 - 323 | 4.47 | 0.19 | 0.71 | 17.85 | 1.57 | 11.75 |
| 2,5-Dimethylhexane | 550.1 | 24.9 | 5.620 | 0.356 | 382.3 | 346.3 | 475.67 | 273 - 373 | 2.74 | 0.13 | 0.82 | 10.10 | 1.19 | 8.49 |
| 3,3-Dimethylhexane | 562. | 26.5 | 5.583 | 0.320 | 385.1 | 343.05 | 458.96 | 273 - 323 | 5.57 | 0.10 | 0.69 | 22.61 | 2.31 | 17.52 |
| 3,4-Dimethylhexane | 568.9 | 26.9 | 5.561 | 0.338 | 390.9 | 342.5 | 461.68 | 273 - 333 | 4.82 | 0.12 | 0.54 | 19.15 | 0.99 | 12.48 |
| 2-Methyl-3-Ethylpentane | 567.1 | 27. | 5.589 | 0.330 | 388.8 | 338.3 | 458.16 | 273 - 323 | 5.47 | 0.16 | 0.40 | 16.90 | 1.08 | 14.18 |
| 3-Methyl-3-Ethylpentane | 576. | 28.1 | 5.542 | 0.303 | 391.4 | 340.1 | 448.01 | 273 - 323 | 7.28 | 0.02 | 0.46 | 30.86 | 2.77 | 22.13 |
| 2,2,4-Trimethylpentane | 544. | 25.7 | 5.620 | 0.303 | 372.4 | 344.3 | 451.63 | 273- 363 | 7.00 | 0.07 | 0.62 | 32.36 | 4.30 | 23.08 |
| 2,2-Dimethylheptane | 576.8 | 23.5 | 6.286 | 0.390 | 405.9 | 373.1 | 513.69 | 283 - 323 | 5.34 | 0.17 | 0.67 | 10.48 | 3.48 | 14.79 |
| 2,2,4-Trimethylhexane | 573.7 | 23.5 | 6.304 | 0.321 | 399.7 | 381.6 | 505.73 | 283 - 333 | 5.91 | 6.31 | 0.21 | 28.53 | 2.80 | 27.53 |
| 2,2,5-Trimethylhexane | 568. | 23.7 | 6.288 | 0.357 | 397.2 | 383.9 | 507.84 | 283 - 333 | 5.84 | 0.02 | 0.19 | 29.34 | 3.26 | 20.81 |
| 3,3-Diethylpentane | 610. | 26.7 | 6.172 | 0.338 | 419.3 | * | 488.40 | 273 - 333 | 8.28 | 0.50 | 0.48 | ** | 3.25 | 26.84 |
| 2,2,3,4-Tetramethylpentane | 592.7 | 26. | 6.279 | 0.313 | 406.1 | 378.65 | 481.35 | 283 - 323 | 10.9 | 25.41 | 0.24 | 52.68 | 0.24 | 33.55 |
| n-Propylcyclopentane | 603. | 30. | 5.307 | 0.335 | 404.1 | * | 450.04 | 273 - 373 | 1.21 | 14.81 | 0.60 | ** | 0.59 | 12.74 |
| Iso-propylcyclopentane | 601. | 30. | 5.297 | 0.240 | 399.6 | * | 446.28 | 273 - 373 | 1.85 | 0.89 | ** | 1.09 | 30.55 | |
| n-Propylcyclohexane | 639. | 28. | 5.965 | 0.258 | 429.6 | 360.4 | 499.48 | 273 - 343 | 2.49 | 2.20 | 0.40 | 9.10 | 2.08 | 35.50 |
| Iso-butylcyclohexane | 659. | 31.2 | 6.635 | 0.319 | 444.5 | 397.7 | 442.55 | 273 - 373 | + | 0.60 | 2.08 | 161.78 | 15.91 | 60.92 |
| Sec-butylcyclohexane | 669. | 26.7 | 6.593 | 0.264 | 452.5 | 397.5 | 533.01 | 273- 373 | + | 1.37 | 1.37 | 24.42 | 1.38 | 48.08 |
| Tert-butylcyclohexane | 659. | 26.6 | 6.599 | 0.252 | 444.7 | 394.65 | 524.52 | 273 - 373 | + | 1.79 | 1.62 | 28.64 | 1.63 | 51.89 |
| 1-Hexene | 504. | 31.7 | 4.181 | 0.285 | 336.6 | * | 437.18 | 273 - 333 | 1.54 | 0.52 | 0.71 | ** | 0.68 | 2.59 |
| 1-Octene | 566.7 | 26.2 | 5.550 | 0.386 | 394.4 | * | 472.71 | 273 - 373 | 0.76 | 0.30 | 0.25 | ** | 0.45 | 2.77 |
| 1-Decene | 615. | 22. | 6.880 | 0.558 | 486.5 | * | 581.81 | 273 - 373 | 1.70 | 0.10 | 0.79 | ** | 0.80 | 0.72 |
| 1-Dodecene | 657. | 318.5 | 8.206 | 0.558 | 486.5 | * | 704.61 | 273 - 373 | 0.29 | 0.60 | 0.94 | ** | 1.97 | 0.54 |
| n-Propylbenzene | 638.2 | 32. | 5.790 | 0.344 | 432.4 | 323.35 | 448.29 | 273 - 373 | 2.01 | 0.25 | 1.04 | 8.14 | 1.05 | 17.58 |
| Iso-propylbenzene (Cumene) | 631.1 | 32.1 | 5.786 | 0.326 | 425.6 | 321.1 | 440.06 | 273 - 373 | 3.19 | 0.16 | 1.22 | 13.07 | 1.22 | 22.16 |
| 2-EthylToluene | 651. | 30.4 | 5.790 | 0.294 | 438.3 | 320.0 | 483.96 | 273 - 373 | 6.50 | 1.51 | 1.10 | 23.41 | 6.11 | 20.17 |
| 3-Ethyltoluene | 637. | 28.4 | 5.816 | 0.360 | 434.5 | 322.15 | 503.41 | 273 - 373 | 9.58 | 7.98 | 1.73 | 32.43 | 6.93 | 6.76 |
| 4-Ethyltoluene | 640. | 29.4 | 5.833 | 0.322 | 435.2 | 323.3 | 486.78 | 273 - 373 | 6.15 | 1.52 | 1.57 | 22.29 | 3.93 | 16.98 |
| n-Butylbenzene | 660.5 | 28.9 | 6.455 | 0.393 | 456.5 | 362.9 | 497.21 | 273 - 373 | 3.08 | 0.04 | 1.74 | 13.95 | 1.89 | 18.91 |
| Iso-Butylbenzene | 650. | 31.4 | 6.471 | 0.380 | 445.9 | 360.3 | 445.77 | 273 - 373 | + | 3.13 | 1.01 | 70.96 | 9.33 | 32.28 |
| Sec-Butylbenzene | 664. | 29.4 | 6.445 | 0.274 | 442.3 | 359.95 | 488.71 | 273 - 373 | + | 2.25 | 1.22 | 18.24 | 1.21 | 45.00 |
| Tert-Butylbenzene | 660. | 29.6 | 6.440 | 0.265 | 442.3 | 356.8 | 481.45 | 273 - 373 | + | 1.78 | 1.31 | 21.10 | 1.39 | 48.27 |
| 1,4-Diethylbenzene | 657.9 | 28. | 6.518 | 0.404 | 456.9 | 361.45 | 509.95 | 273 - 373 | + | 0.31 | 1.44 | 1.28 | 1.45 | 15.53 |
| 1-Methylnaphthalene | 772. | 36. | 6.980 | 0.310 | 517.9 | 353.8 | 483.33 | 273 - 303 | 3.81 | 0.11 | 0.60 | 17.63 | 5.39 | 44.00 |
| 1-Tetradecene | 689. | 15.6 | 9.471 | 0.644 | 524.3 | * | 839.84 | 293 - 373 | 2.12 | 8.15 | 0.94 | ** | 3.31 | 4.94 |
| Carbon Tetrachloride | 556.4 | 49.6 | 3.784 | 0.193 | 349.9 | 219.68 | 279.37 | 288 - 525 | + | 8.95 | 2.62 | 42.64 | 2.73 | 15.50 |
| Chloroform | 536.4 | 53.7 | 3.071 | 0.218 | 334.3 | 183.4 | 260.02 | 288 - 348 | + | 1.88 | 0.76 | 3.29 | 1.98 | 1.49 |
| 1,1,2-Tricloroethane | 606. | 51.4 | 3.701 | 0.2598 | 386.7 | 223.8 | 304.39 | 288 - 378 | + | 5.76 | 0.65 | 15.34 | 4.57 | 0.84 |
| 1-Chlorobutane | 542. | 36.8 | 3.641 | 0.218 | 351.6 | 230.3 | 359.96 | 283 - 343 | + | 3.23 | 0.41 | 34.34 | 5.96 | 1.68 |
| Fluorobenzene | 560.1 | 45.5 | 3.742 | 0.244 | 357.9 | 214.15 | 306.62 | 283 - 353 | + | 0.07 | 0.21 | 5.92 | 1.40 | 6.24 |
| Chlorobenzene | 632.4 | 45.2 | 4.458 | 0.249 | 404.9 | 244.4 | 344.11 | 283 - 403 | + | 0.05 | 0.69 | 0.84 | 2.85 | 15.98 |
| Bromobenzene | 670. | 45.2 | 4.858 | 0.251 | 429.2 | 258.32 | 359.26 | 283 - 423 | + | 0.11 | 1.20 | 6.25 | 1.34 | 25.58 |
| Iodobenzene | 721. | 45.2 | 5.602 | 0.249 | 461.6 | 279.19 | 378.58 | 283 - 433 | + | 0.11 | 1.26 | 18.48 | 1.66 | 41.41 |
| Acetone | 508.1 | 47. | 2.316 | 0.304 | 329.2 | 161.22 | 297.03 | 293 - 333 | + | 0.27 | 2.71 | 66.20 | 18.89 | 35.07 |
| Benzonitrile | 699.4 | 42.2 | 4.500 | 0.362 | 464.3 | * | 414.43 | 293 - 363 | + | 0.21 | 0.53 | ** | 6.70 | 6.072 |
| Dibutylether | 580. | 25.3 | 5.866 | 0.502 | 413.4 | * | 493.37 | 283 - 393 | + | 2.37 | 0.11 | ** | 0.22 | 12.64 |
| Cyclooctane | 647.2 | 35.6 | 5.255 | 0.236 | 422.0 | 315.15 | 417.11 | 283 - 393 | + | 0.26 | 1.82 | 29.38 | 1.87 | 34.57 |
| **OVERALL AAD%** | | | | | | | | | **2.80** | **1.76** | **1.05** | **16.81** | **2.57** | **16.75** |

(*) The experimental parachor for these substances was not available
(+) Experimental densities for these compounds are not readily available
(**) These values could not be calculated since there are no experimental parachor data available.
Experimental densities were taken from Hall, 1986; experimental surface tensions from Jasper (1972), Hall (1986), Grigiryev *et al.* (1992), Beaton and Hewitt (1989); the physical properties Tc, Pc, ω, and Tb from Reid *et al.* (1988). $\mathcal{R}^* = \mathcal{R}m/\mathcal{R}m,ref$, $\mathcal{R}m$ data were taken from Hall (1986) or calculated using the Lorentz-Lorenz function, $\mathcal{R}m,ref$ = 6.987 for the reference fluid (methane). Experimental parachor data were taken from Quayle (1953).

Table 2. Comparison of experimental data and predicted values of surface tension with other methods for several representative compounds (*i.e.* low-boiling-point, linear and branched alkanes; cyclic and branched cyclic; aromatic and substituted aromatic; as well as halogenated compounds). The AAD% reported here refers to $|(\sigma_{exp} - \sigma_{calc})/ \sigma_{exp} .100|$.

| Compound | Temperature (°K) | Macleod (1923) | Brock and Bird (1955) | Sivaraman et al (1984) | This work $\mathcal{P}^b$ and Eq10 | This work Eqs10&14 |
|---|---|---|---|---|---|---|
| Methane | 93.15 | | 12.11 | 6.37* | 0.13 | 1.22 |
|  | 103.15 | | 12.31 | 8.97* | 0.57 | 1.96 |
|  | 113.15 | | 12.81 | 11.09* | 0.47 | 0.87 |
| n-Butane | 203.15 | 11. | 1.6 | 6.24* | 0.18 | 1.53 |
|  | 233.15 | 5.2 | 0.9 | 6.67* | 0.86 | 0.86 |
| n-Heptane | 293.15 | 0.6 | 0.3 | 0.89** | 0.05 | 0.15 |
|  | 313.15 | 0.7 | 0.1 | 1.43** | 0.11 | 0.22 |
|  | 323.15 | 3.1 | 0.3 | 1.85** | 0.02 | 0.04 |
| n-Nonane | 298.15 | 1.35 | 0.35 | 0.58** | 0.31 | 1.18 |
| 2-Methyl-3-Ethylpentane | 273.15 | | 0.69 | 14.65* | 0.08 | 0.90 |
|  | 293.15 | | 0.88 | 15.06* | 0.32 | 1.30 |
|  | 313.15 | | 1.05 | 15.52* | 0.51 | 1.47 |
|  | 333.15 | | 1.3 | 16.04* | 0.22 | 1.23 |
|  | 353.15 | | 1.45 | 16.50* | 0.31 | 0.68 |
|  | 373.15 | | 1.51 | 16.98* | 1.12 | 0.15 |
| Cyclopentane | 293.15 | 5.6 | 2.0 | 1.46* | 0.45 | 1.72 |
|  | 313.15 | 2.4 | 0.1 | 3.93* | 1.26 | 0.25 |
| Cyclohexane | 293.15 | 3.9 | 4.8 | 11.44* | 0.13 | 2.71 |
|  | 313.15 | 3.5 | 4.5 | 11.15* | 0.22 | 2.40 |
|  | 333.15 | 2.2 | 4.3 | 10.64* | 0.33 | 2.25 |
| Ethylcyclopentane | 273.15 | | 0.05 | 9.15* | 0.19 | 0.15 |
|  | 293.15 | | 0.47 | 9.34* | 0.07 | 0.44 |
|  | 313.15 | | 0.72 | 7.96* | 0.23 | 0.59 |
|  | 343.15 | | 1.48 | 10.47* | 0.69 | 0.32 |
| Benzene | 293.15 | 5.1 | 2.0 | 2.71** | 0.49 | 3.5 |
|  | 313.15 | 5.6 | 1.9 | 2.94** | 0.19 | 3.2 |
|  | 333.15 | 5.0 | 0.4 | 3.29** | 0.13 | 3.16 |
| Toluene | 293.15 | | 0.27 | 0.35** | 0.56 | 1.86 |
|  | 313.15 | | 0.035 | 0.05** | 0.26 | 1.07 |
|  | 333.15 | | 5.56 | 0.56** | 1.15 | 0.18 |
| n-Propyl benzene | 313.15 | 0.8 | 0.7 | 2.77* | 0.26 | 0.82 |
|  | 333.15 | 2.1 | 0.1 | 2.88* | 0.48 | 0.16 |
|  | 353.15 | 3.9 | 0.7 | 2.84* | 1.19 | 0.52 |
|  | 373.15 | 6.6 | 1.9 | 2.90* | 1.49 | 0.87 |
| Carbon Tetrachloride | 288.15 | 1.1 | 4.9 | 48.49* | 0.32 | 2.24 |
|  | 308.15 | 1.2 | 5.1 | 48.40* | 0.51 | 2.41 |
|  | 328.15 | 1.0 | 5.0 | 48.28* | 0.35 | 2.30 |
|  | 348.15 | 0.1 | 4.9 | 48.10* | 0.10 | 1.82 |
|  | 368.15 | 2.5 | 4.4 | 47.78* | 0.78 | 1.11 |

(*) These values were calculated using the method outlined by Sivaram *et al.* using the acentric factor, critical pressure and temperature reported in this paper. (**) These values were calculated using the Sivaram *et al.*'s method but using the values for the acentric factor, critical pressure and temperature reported in their paper. These data were found to be slightly different from the ones reported here. The experimental data for all compounds were taken from Jasper (1972).

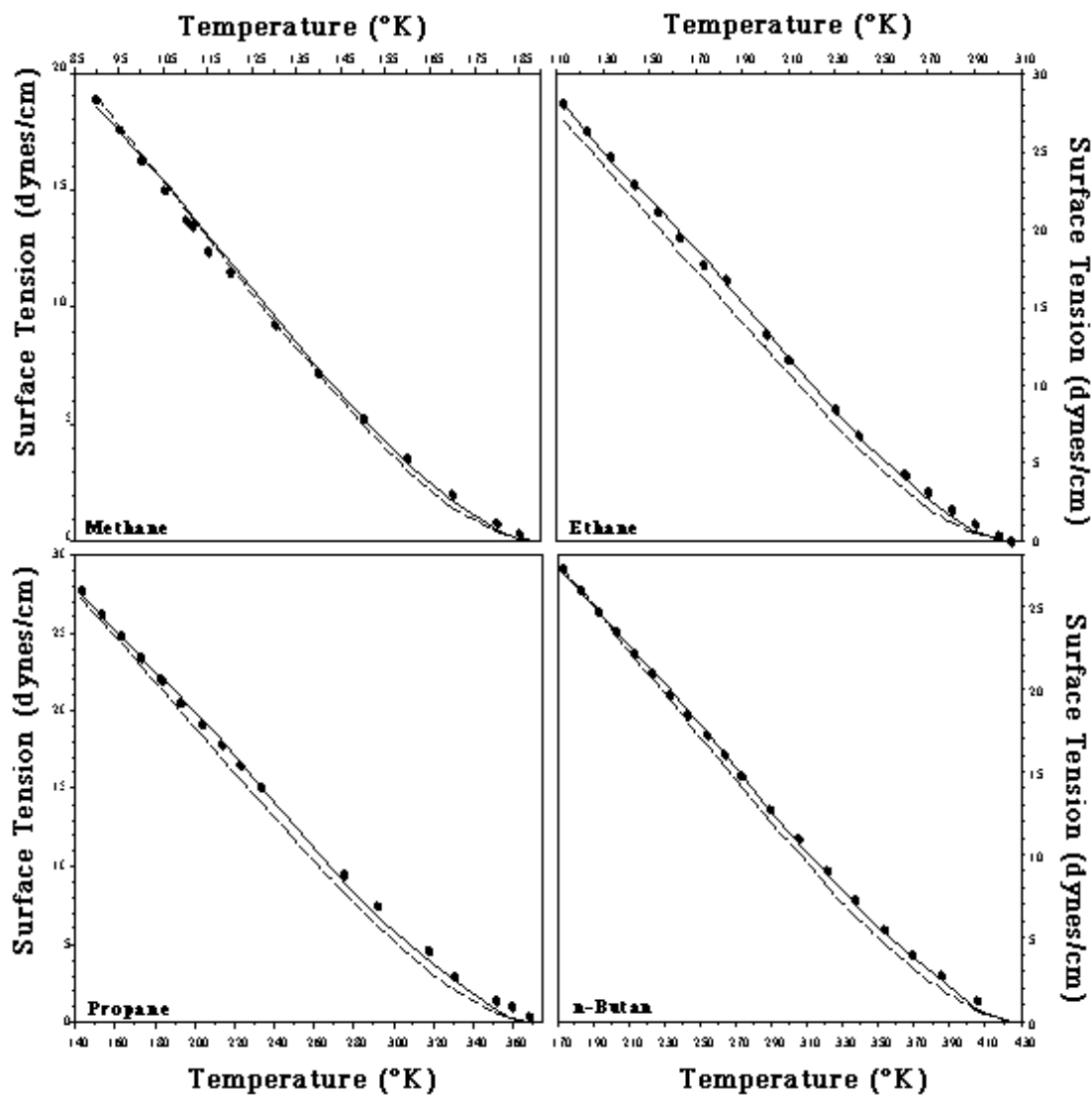

Figure 1. Surface Tension, s, as a function of temperature for low-boiling-point compounds. The filled circles indicate the experimental data (Jasper, 1972; Hall, 1986; Beaton and Hewitt, 1989). The solid line represent the values calculated with Equation 10, whereas the broken line represents the values calculated using the experimental parachor along with Equation 2.

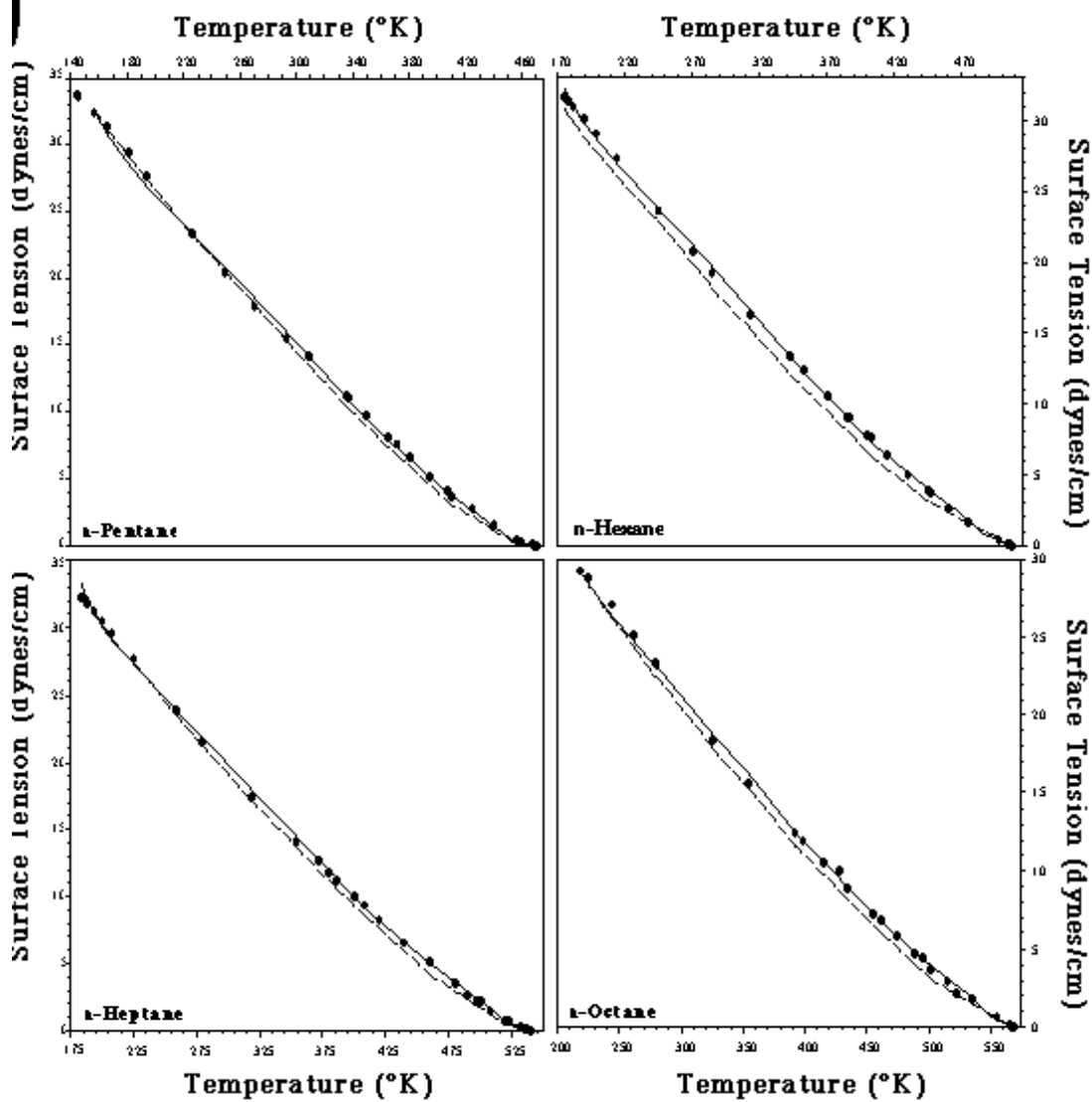

Figure 2. Surface Tension, s, of normal paraffins as a function of temperature from the triple point to the critical point. The filled circles indicate the experimental data (Grigoryev, et al., 1992) The solid line represents the values calculated with Equation 10, whereas the broken line represents the values calculated using the experimental parachor along with Equation 2.

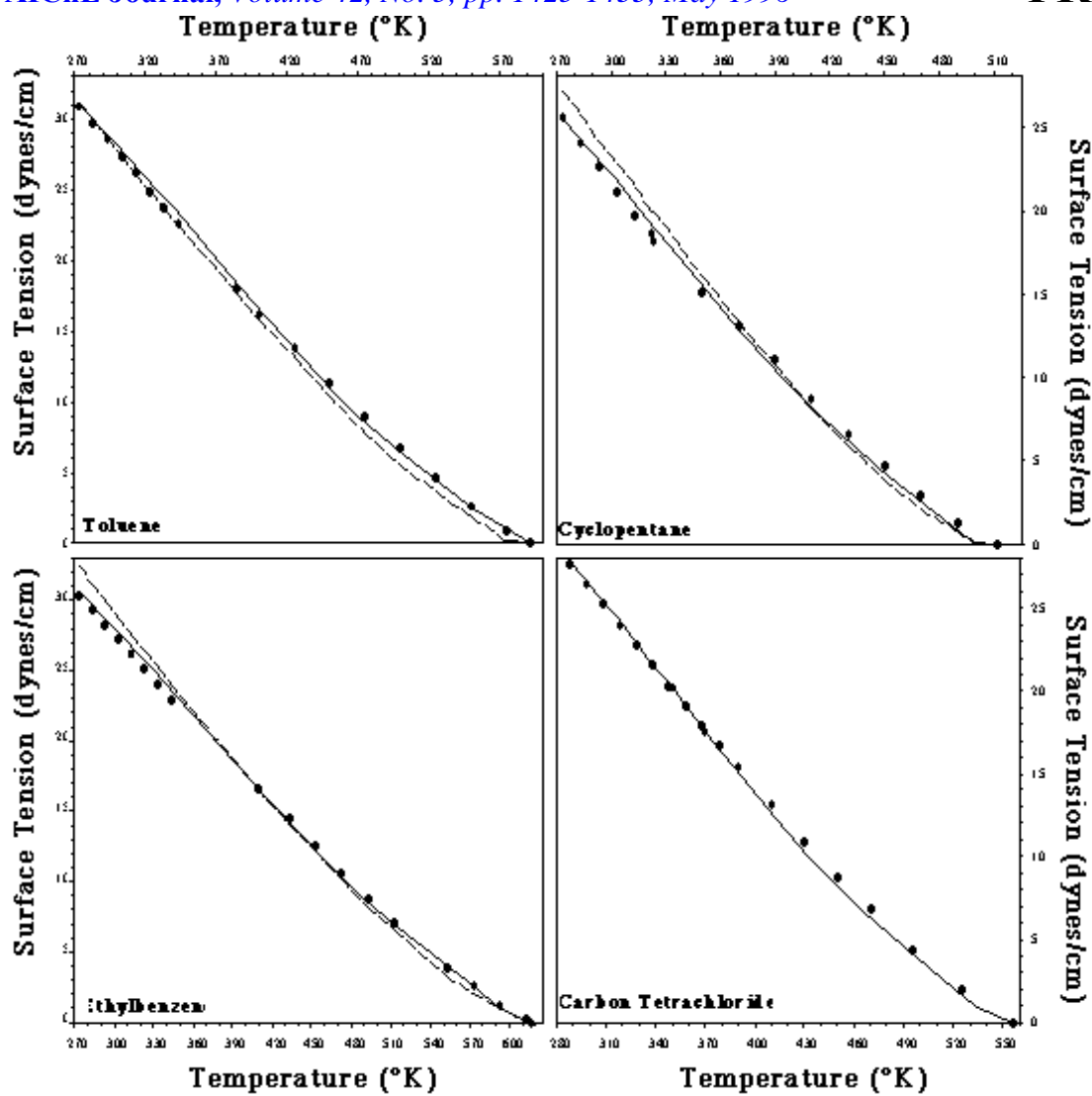

Figure 3. Surface Tension, s, as a function of temperature for various compounds. The filled circles indicate the experimental data (Jasper, 1972; Hall, 1986; Beaton and Hewitt, 1989). The solid line represent the values calculated with Equations 10, whereas the broken line represents the values calculated using the experimental parachor along with Equations 2.

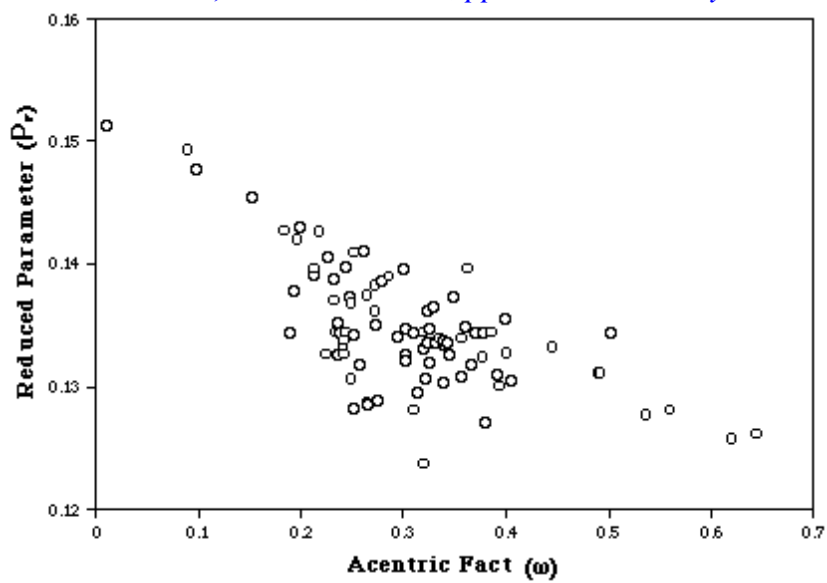

Figure 4. Reduced parameter ($P_r$) versus acentric factor ($\omega$).

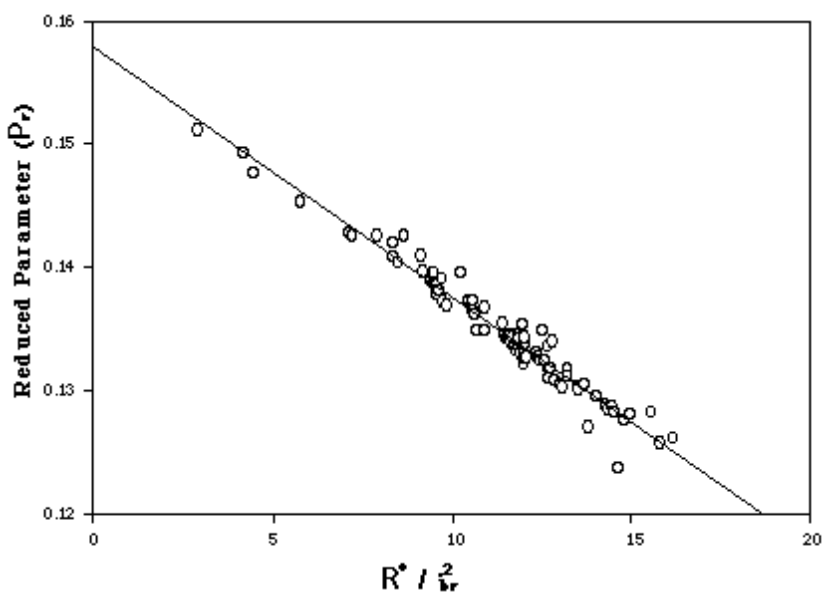

Figure 5. Reduced Parameter ($P_r$) versus ($R^*/T_{br}^2$), $R^*$ is the dimensionless molar refraction defined as $R^* = R_m/R_{m,ref}$, $R_m$ is the molar refraction and $R_{m,ref}$ is the molar refraction of the reference substance which in our case is methane with a value of 6.987. $T_{br}$ is the reduced normal-boiling-point.